\documentclass[aps,prb,twocolumn,superscriptaddress, show pacs]{revtex4-1}
\usepackage{bm, amsmath}
\usepackage{graphicx}
\usepackage{color}
\usepackage{epstopdf}
\usepackage{graphicx}
\usepackage{dcolumn}
\usepackage{bm}
\usepackage{subfigure}

\begin{document}

\title{{Spintronics via non-axisymmetric chiral skyrmions}}


\author{A.~O.~Leonov}
\thanks{A.Leonov@ifw-dresden.de}
\affiliation{Center for Chiral Science, Hiroshima University, Higashi-Hiroshima, 
Hiroshima 739-8526, Japan}
\affiliation{IFW Dresden, Postfach 270016, D-01171 Dresden, Germany}   
\affiliation{Zernike Institute for Advanced Materials, University of Groningen, 
Groningen, 9700AB, The Netherlands}

\author{J.~C.~Loudon}
\affiliation{Department of Materials Science and
  Metallurgy, 27 Charles Babbage Road, Cambridge, CB3 0FS, United Kingdom}

\author{A.~N.~Bogdanov}
\affiliation{Center for Chiral Science, Hiroshima University, Higashi-Hiroshima, 
Hiroshima 739-8526, Japan}
\affiliation{IFW Dresden, Postfach 270016, D-01171 Dresden, Germany}

\date{\today}

\begin{abstract}
{ Micromagnetic calculations demonstrate a peculiar evolution of
  non-axisymmetric skyrmions  driven by an applied 
	magnetic field in confined helimagnets with longitudinal
  modulations. We argue that these {specific} 
	solitonic states can be employed in nanoelectronic devices as 
	an effective alternative to the common axisymmetric skyrmions  
	which occur in magnetically saturated states.  }
\end{abstract}

         
\maketitle

Two-dimensional topological solitons with an axisymmetric structure 
(commonly addressed as  isolated  \textit{chiral skyrmions} \cite{Leonov16a}) 
are stabilized by Dzyaloshinskii-Moriya interactions in the \textit{saturated} 
states of noncentrosymmetric magnetic materials \cite{Bogdanov94}.
In magnetic nanolayers, chiral skyrmions represent nanosized spots of
reverse magnetization which can be created or deleted by
a magnetic tip \cite{Romming13} and moved by electric currents and
applied magnetic fields \cite{Iwasaki2013,Romming13,Leonov16a}.
Due to their remarkable properties, magnetic skyrmions are considered promising objects for next-generation memory and logic devices \cite{Kiselev11,Fert2013,Tomasello2014}, 
%
which store information in the form of skyrmions  that can be manipulated at room temperature \cite{Moreau2015,Woo2016,Yu2012}.

\begin{figure}
\includegraphics[width=0.97\columnwidth]{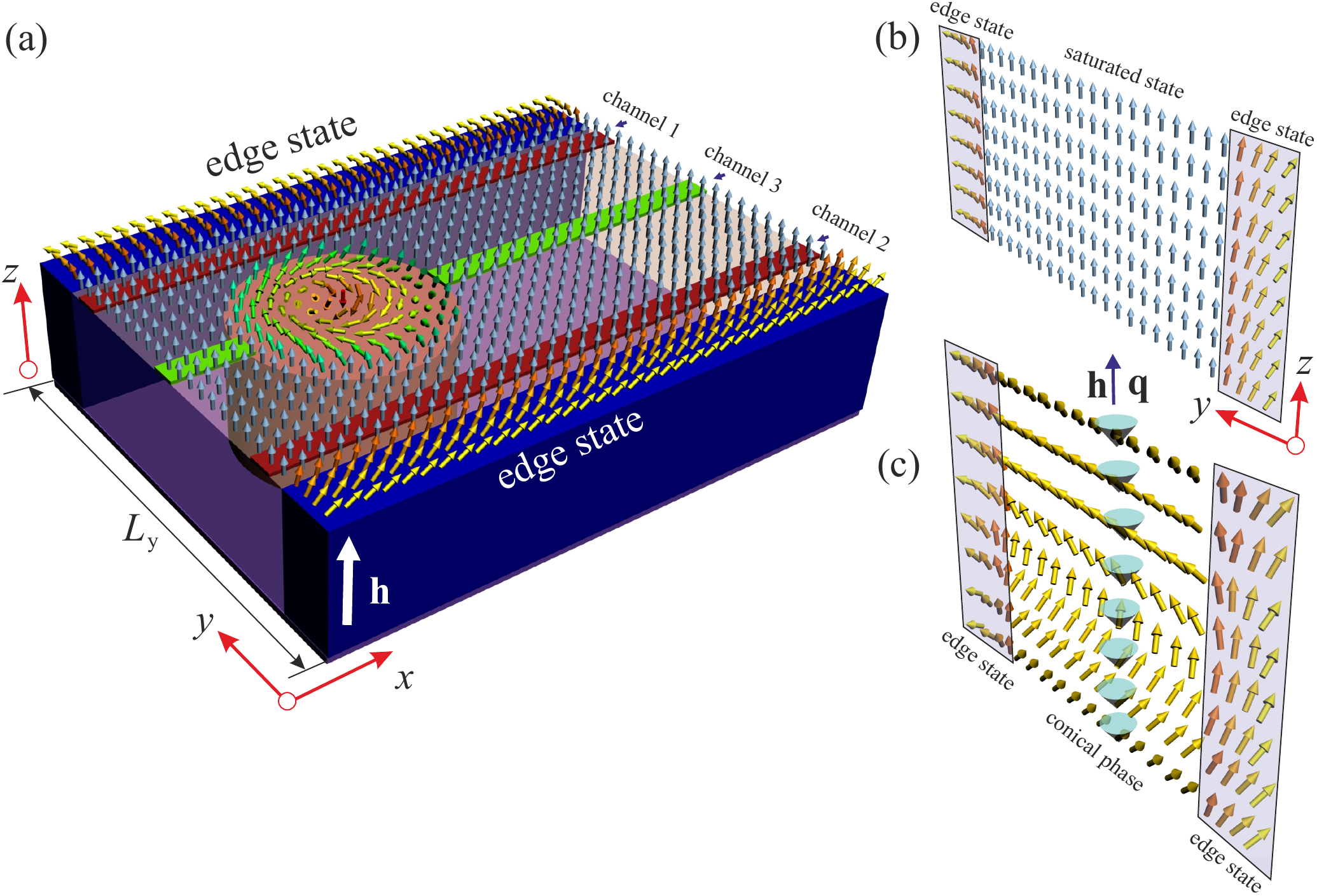}
\caption{ 
(color online). (a) Schematic of the motion of an isolated skyrmion in a film  
infinite in the $x$ and $z$ directions (periodic boundary conditions with 
period $p$ are used for these coordinates) and confined by parallel planes 
at $y=0; L_y$. 
At high magnetic fields with $H$ parallel to $z$
($H>H_D$), axisymmetric skyrmions exist within the
magnetically saturated matrix.  Repulsive edge modulations force
axisymmetric skyrmions to locate along the middle line of
the sample (indicated with the \textit{green} line). For $H<H_D$ the
saturated state transforms into the longitudinally modulated
(\textit{cone}) phase with the propagation vector $\mathbf{q}$ along the field, and skyrmions become non-axisymmetric and
inhomogeneous along the thickness (see model (\ref{model})). In this
case, under the influence of attractive interactions with the edge
states, the skyrmions are situated along facets of the sample
(indicated by two red strips). (b) and (c) schematically show the structure of the homogeneous and the conical phases in the $yz$ cross section, correspondingly.
\label{art}
}
\end{figure}

In practice, isolated magnetic skyrmions are induced and manipulated in laterally confined saturated helimagnets (slabs, narrow strips, nanowires, and nanodots) \cite{Romming13, Leonov16a,Hanneken16,Du,Tomasello2014}. 
%
%
Importantly, magnetic saturation is never fully reached in confined nanosystems as surface modulations occur near the sample
edges (so called \textit{chiral surface twists}) \cite{twist} with a \textit{penetration depth} 
estimated as 0.1 $p$ ($p$ is the helix period at zero field)  \cite{Meynell14}. 
%
In the case of a narrow strip, the edge states manifest themselves as remnants of the helical spiral \cite{Wilson2013,Keesman2015,Meynell14} with a smooth deviation of the magnetization from being co-aligned with the field in the middle of the sample (Fig. \ref{art} (a), (b)) to composing some (field- and anisotropy-dependent) angle $\theta_0$  at the edge. 
%
%
To date, theoretical investigations  of confined chiral skyrmions and their applications   have been restricted  to saturated helimagnets (Fig. \ref{art}(b))   \cite{Fert2013,Zhang2015}.
 In that case, the skyrmion-edge interaction has a
repulsive character (Eq. (21) in Ref. \onlinecite{Meynell14}) due to
the same rotational sense of the magnetization in the
\textit{axisymmetric} skyrmions and the surface modulations.

In this Letter we address a {special} type of \textit{non-axisymmetric} 
skyrmion introduced in Ref. \onlinecite{Leonov16b}.
These three-dimensional solitonic states arise in longitudinally modulated chiral ferromagnets 
(with the conical phase, Fig. \ref{art} (c))  and hence are inhomogeneous along their axes.
Within the micromagnetic model we calculate the structure of  non-axisymmetric skyrmions 
and  edge modulations in a confined chiral helimagnet. 
We show that the conical phase turns the skyrmion-edge repulsion into an attraction and consequently, 
there is an equilibrium distance from the edge at which the force on the skyrmions is zero. 
This equilibrium distance can be tuned by changing the applied
magnetic field and it acts to guide the skyrmions along the edges.
We demonstrate that specific properties of confined non-axisymmetric skyrmions 
offer new directions in spintronic applications of chiral skyrmions.

The outline of this article is as following: first, we investigate the structure of the edge states arising in the conical phase at the lateral boundaries of the system (Fig. 2). We do not include isolated skyrmions at this stage. Then, we proceed with a brief theoretical
overview of isolated skyrmions within the conical phase without any influence of the edge states (Fig. 3). Based on this, we finally consider edge-skyrmion attraction in Fig. 4.

%

The equilibrium solutions for skyrmions and edge modulations are derived  
within the standard discrete model of a chiral ferromagnet  where the total 
energy is given by:
%
\begin{align}
&w =  J\,\sum_{<i,j>} (\mathbf{S}_i \cdot \mathbf{S}_j ) -\sum_{i} \mathbf{H} \cdot \mathbf{S}_i - D \, \sum_{i} (\mathbf{S}_i \times \mathbf{S}_{i+\hat{x}} \cdot \hat{x} \nonumber\\
& + \mathbf{S}_i \times \mathbf{S}_{i+\hat{y}} \cdot \hat{y} +\mathbf{S}_i \times \mathbf{S}_{i+\hat{z}} \cdot \hat{z}). 
 \label{model}
\end{align}
$S_i$ is the unit vector in the direction of the magnetization at the site $i$ of a three-dimensional cubic
lattice and $<i,j>$ denote pairs of nearest-neighbor spins.
The first term describes the ferromagnetic nearest-neighbor exchange with $J<0$, the second term  is the Zeeman interaction, and the third term stands for the Dzyaloshinskii-Moriya (DM) interaction. 
The DM constant $D = J \tan (2\pi/p)$ defines the period of modulated structures $p$.
{It was established by direct calculations
that in chiral ferromagnets the DM interactions 
strongly suppress demagnetization effects \cite{Bogdanov94,Kiselev11}.
In many practical cases the surface and internal stray-field magnetostatic 
energy of skyrmions can be reduced to local energy contributions and 
included into effective magnetic anisotropy energy
\cite{Bogdanov94,Kiselev11}.}

In what follows, we use $J=1$ and the DM constant is set to $0.445$ which  corresponds to $p=15$. 
We consider  periodic boundary conditions in $z$ and $x$ directions, whereas along $y$  the stripe is confined by vertical surfaces with the  free boundary conditions (Fig. \ref{art}). The size of our numerical grid is set to $2p\times 50\times p$. 
In an infinite sample,  
below the critical field $H_D=D^2/2J$, the global minimum of (\ref{model}) corresponds to the modulation phase with the propagation direction along the applied field,
the \textit{cone} phase \cite{Bak80} (Fig. \ref{art} (c)):
\begin{equation}
 \theta_c=\arccos(H/H_D), \quad \quad \psi_c=2\pi z/p, 
\label{cone}
\end{equation}
where $\theta, \psi$ are the polar and azimuthal angles of the magnetization vector. 

\begin{figure}
\includegraphics[width=0.9\columnwidth]{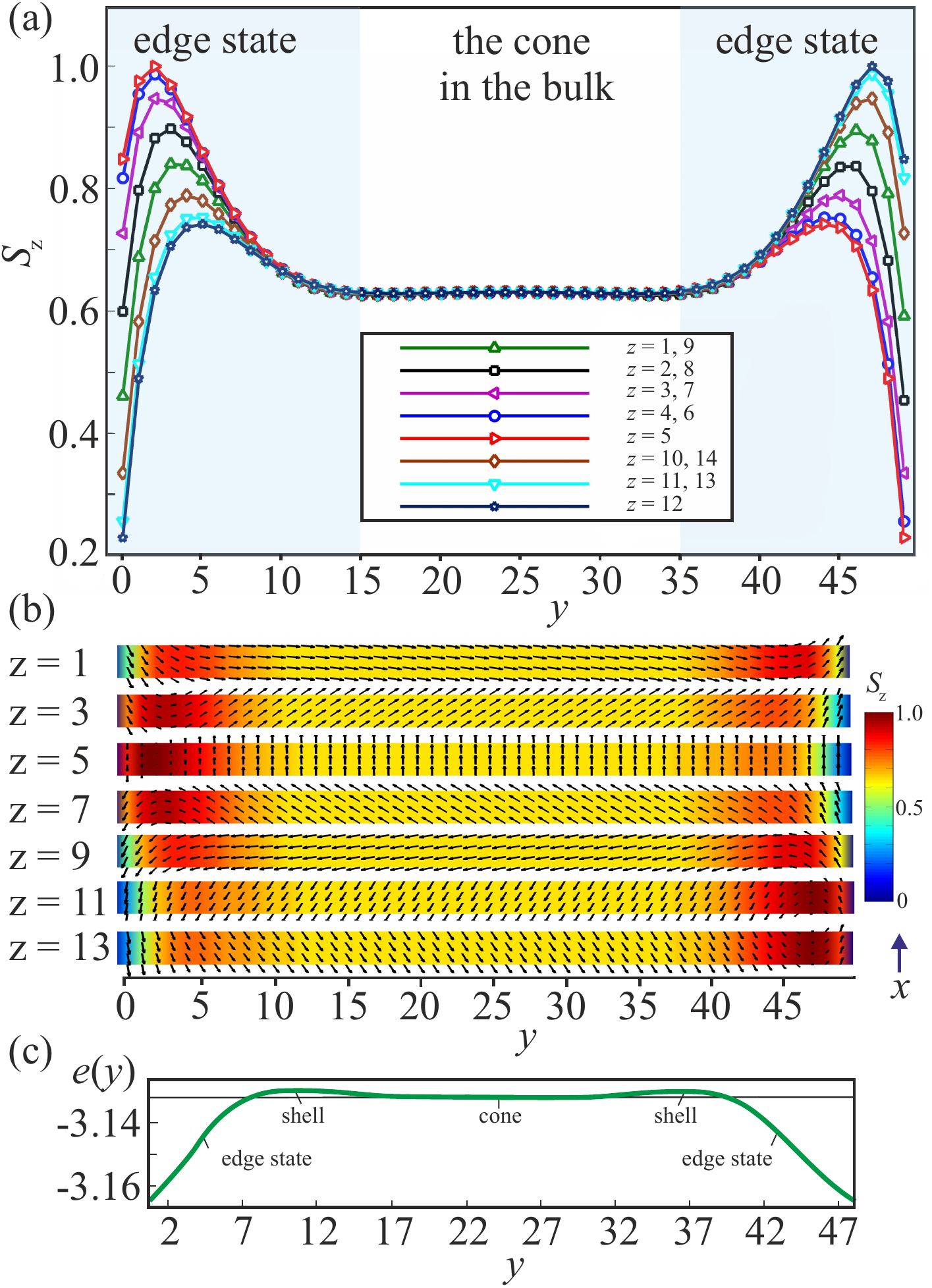}
\caption{ 
(color online) Edge states in a film of a chiral magnet with the conical phase (\ref{cone}). (a) Variations of the magnetization component $S_z$ along the $y$ coordinate for fixed values of $z$. The blue shading indicates the regions occupied by the edge states. 
(b) color plots of $S_z(x; y)$ in a $xy$ plane with fixed values of $\psi_c$ and $z$. The black arrows show in-plane spin-components.  
(c) The energy density $e(y)$ averaged over $z$ (green solid line). The black thin line shows the energy density in the conical phase. The energy density is divided into constituent parts: the energy distribution near the edges, in the conical phase and in the shell which is formed due to the incompatibility of the spin structures at the edges and the conical phase in the middle of the sample.
\label{edge}
}
\end{figure}

\begin{figure}
\includegraphics[width=0.85\columnwidth]{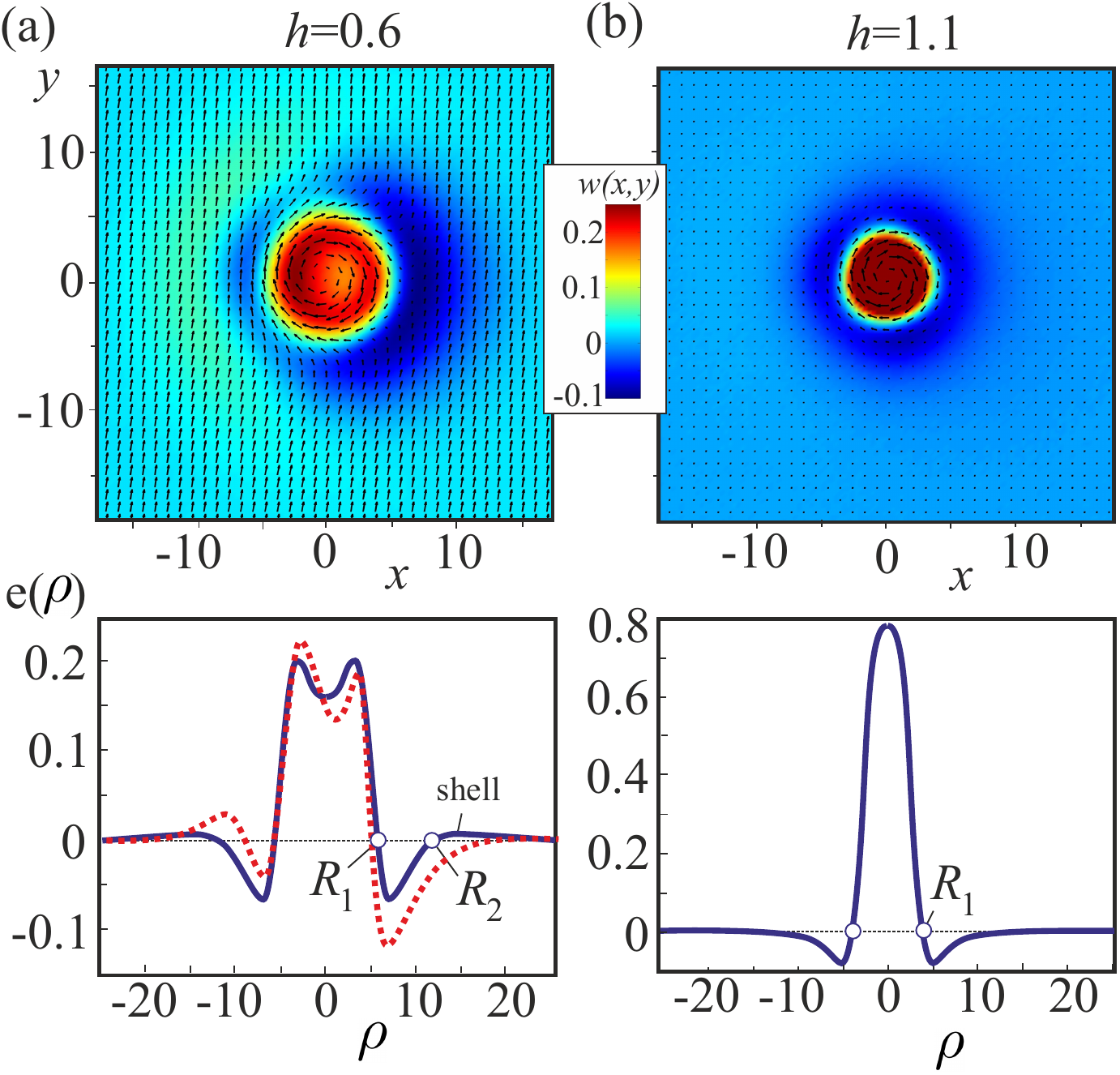}
\caption{ 
(color online) Numerical solutions for isolated skyrmions obtained within the continuum version of the model (\ref{model}). In (a) and (b) the isolated skyrmions are placed within different backgrounds - within the conical phase for $H/H_D=0.6$ and the saturated state for $H/H_D=1.1$, respectively. The first horizontal row shows two color plots of the energy density distributions $w(x;y;z=\mathrm{const})$. The energy density is measured in units $A/D^2$ where $A$ and $D$ are the constants of the exchange ($A\,(\mathbf{grad}\,\mathbf{m})^2$) and DM ($D\,\mathbf{m}\cdot \mathrm{rot}\,\mathbf{m}$) interactions in the continuum version of (\ref{model}).  In the second row the energy density $e(\rho)$ is averaged over the $z$-coordinate and plotted across the skyrmionic centers. The red dotted line shows the cross-section of the color plot $w(x;y;z=\mathrm{const})$.
The characteristic radius $R_2$ signifies the formation of a skyrmionic shell with positive energy over the conical phase (see also Supplementary Movie). 
\label{skyrmion}
}
\end{figure}

The saturated state with $\theta = 0$  occurs when $ H > H_D$.
In the saturated state, isolated skyrmions are axisymmetric and translationally invariant along $z$ \cite{Bogdanov94,Leonov16a}. 
Chiral surface twists for $H > H_D$ have been investigated in a number of earlier contributions \cite{Wilson2013,Meynell14,Keesman2015}.
Below $H_D$, the incompatibility with the longitudially modulated conical phase imposes a complex, three-dimensional character on the magnetic modulations of the skyrmions and edge states (Figs. \ref{edge}-\ref{skyrmion}).


The solutions for  edge states in the sample with the conical phase (\ref{cone}) in its bulk are shown in Fig. \ref{edge}.
The dependence $S_z(y)$ for a fixed value of $x$ and $z$ (Fig. \ref{edge} (a)) shows the formation of two humps in the vicinity of two free surfaces $xz$ with $y=0$ and $y=L_y$. The $S_z$-component in these humps is larger than that for the cone in the middle of the sample and for some cross-sections ($z=5, 12$) it even goes through the state co-aligned with the field, $S_z=1$ (shaded regions in Fig. \ref{edge} (a) and the color plots in Fig. \ref{edge} (b)). 
As the edge states are incompatible with the conical phase, they are surrounded by a strip-like ``shell" - a transitional region running parallel to the boundary.
The shell has the higher energy density
  as compared with the conical phase (Fig. \ref{edge} (c)) and positive exponentially decaying asymptotics.

\begin{figure}
\includegraphics[width=0.9\columnwidth]{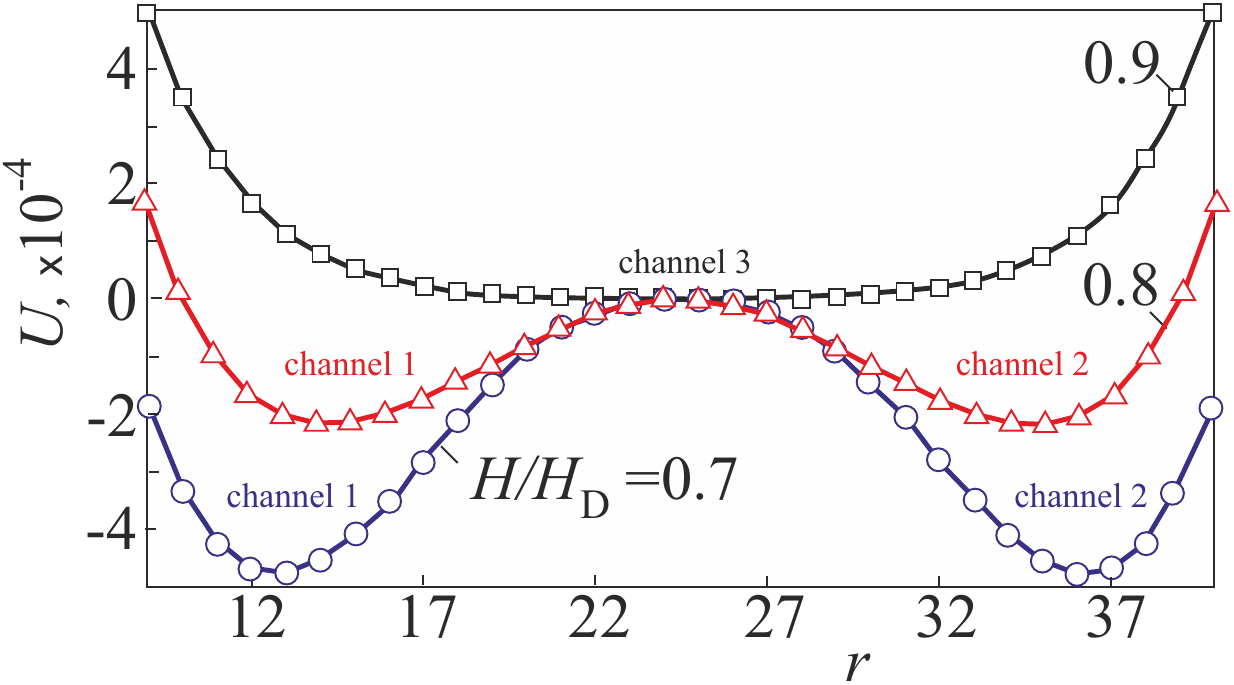}
\caption{ 
(color online). The skyrmion-edge potential energy $U$ vs. the distance $r$ between the skyrmion center and the edge of the sample. The potential energy is measured with respect to the energy of a system for $r=L_y/2$, which corresponds to a skyrmion in the middle of the sample. $U(r)$ was calculated by imposing the constraint, $S_z = -1$, at the skyrmion center and minimizing the energy with respect to spins at all other sites. The local minima of $U(r)$ give rise to a sequence of edge channels: channels numbered 1 and 2 are located at the boundaries of the sample for $H<H_D$, and the channel 3 runs along the sample center for $H>H_D$. These channels, separated by the potential barriers, guide the  motion of skyrmions. 
%
\label{attraction}
}
\end{figure}

Fig. \ref{skyrmion} shows color plots of the energy density $w(x;y;z=\mathrm{const})$  and  skyrmion energy densities averaged over the $z$-coordinate, $e(\rho)=(1/p)\int_0^p w dz$, and  plotted along the radial directions for isolated skyrmions within the conical phase ($H/H_D=0.6$, Fig. \ref{skyrmion} (a)) and within the saturated state ($H/H_D=1.1$, Fig. \ref{skyrmion} (b)). 
%
For $H>H_D$,  a characteristic radius $R_1$  specifies the size of the skyrmionic core. The core ($\rho<R_1$) with the positive energy density is surrounded by the ring ($\rho>R_1$) with the negative energy density  which is known to form due to the DM interaction and protects isolated skyrmions from collapse \cite{Leonov16a}.
%
The ring has the radial symmetry in all layers with the fixed $z$-coordinate (the color plot of $w(x;y;z=\mathrm{const})$ in Fig. \ref{skyrmion} (b)).
For $H<H_D$, $e(\rho)$ has two characteristic radii $R_1$ and  $R_2$. The skyrmionic shell is the part of the non-axisymmetric skyrmion with $\rho > R_2$ and represents an outer ring \cite{Leonov16b} with the positive energy density. 
The color plot in Fig. \ref{skyrmion} (a)  also shows that the ring with the negative energy density is partially weakened. 
This occurs along the radial directions where the magnetization rotates from the state opposite to the field in the center (polar angle of the magnetization $\theta=\pi$) directly to $\theta_c$. 
On the contrary, the ring is restored in those parts where the magnetization rotation goes to $\theta_c$ via the state with $\theta=0$. 
The attraction between skyrmions and the edges of the
  track occurs because the total energy can be reduced if their
  respective shells overlap.
The skyrmion-edge interaction potentials as a function of a distance $r$ between the skyrmion center and the $(y=0)$-edge of the sample for different values of the applied magnetic field are plotted in Fig. \ref{attraction}. 
Potential profiles show that the attractive skyrmion-edge coupling is characterized by a rather deep potential well
establishing the equilibrium separation of skyrmions from the edges. 
%
%
The distance $r_{min}=\min(r)$ increases rapidly with the field. And already for $H/H_D=0.9$ skyrmions are ``released" by the edge states and are pushed  to the center of the sample. 
%

%

%
Our results open completely new perspectives on using skyrmions in nanoelectronic devices with the conical phase. First of all, the edge states with the complex spin structure give rise to a formation of two edge ``channels" (schematically shown by red strips in Fig. \ref{art}) which run along boundaries of chiral magnets and guide the skyrmions. These channels correspond to  the minima of skyrmion-edge interaction potentials (Fig. \ref{attraction}). 
The distance of channels from  boundaries effectively depends on the value of the applied magnetic field:  for $H/H_D=0.8$ (red line with triangular markers in Fig. \ref{attraction}) the channels are located farther from the boundaries and closer to each other than for $H/H_D=0.7$ (blue line with circular markers in Fig. \ref{attraction}). 
For some threshold field $H_{tr}$,  two channels numbered 1 and 2 in Figs. \ref{art} and \ref{attraction}
 merge into one channel 3 along the  middle of the sample. 
The threshold field of this phenomenon depends mainly on the confinement ratio $\nu$ -- the ratio of $p$ to the width of the racetrack $L_y$. 
Due to the field-dependent position of channels, channels
  1 and 2 may overlap and form the channel 3 for the fields lower than
  $H_D$ in narrow films.
In particular, the sample used in the numerical simulations ($\nu=L_y/p=49/15=3.27$) exhibits one central channel already for $H/H_D=0.9$ in spite of the stable conical phase.
$H_{tr}$ gradually increases for wider films  and reaches $H_D$ for infinitely wide samples.

We also note that for samples confined by parallel surfaces along $z$-axis with $z=0; L_z$,  chiral surface twists additionally modify the structure of skyrmions near the surfaces \cite{Leonov16c}. 
These surface twists distort the translational invariance of skyrmions even in the saturated state and become evident in an additional twist of the azimuthal angle of the magnetization in skyrmions. 
We argue, however, that this fact does not change the attractive nature of the skyrmion-edge potential and subsequent effects.  


%

The channel management by the applied magnetic field opens new ways to do logical operations with skyrmions on racetracks, as the information can be encoded in the lateral positions of skyrmions.
Skyrmions, which fit perfectly into the edge channels, can be directed by currents along two lateral boundaries of a film and may be switched between channels by current pulses. 
As the potential barrier between two lateral channels is lower for larger values of the field, the lower current densities are needed to switch skyrmions between two channels.
On the contrary, to ensure that the skyrmions do not jump from one channel to another due to the skyrmion Hall effect, the magnetic field must be decreased leading to the higher potential barrier between channels.
Moreover, the consecutive  order of skyrmions is also influenced by the value of the field, as isolated skyrmions within the conical phase attract each other and form clusters with the field-dependent inter-skyrmion distance (see for details Ref. \onlinecite{Leonov16b}). This  may help to avoid clogging of skyrmionic bits as encountered in Ref. \onlinecite{Zhang2015}.

To summarize, our data have clearly demonstrated that the skyrmion-edge attraction develops in the presence of the longitudinally modulated phases and may play an important role in skyrmion-based spintronic devices (e.g. a racetrack memory design). 
In particular, it can be employed for magnetic patterning of nanodevices.
The mechanism of skyrmion-edge attraction stems from the complex spin structures of the edge states formed at the boundaries of confined helimagnets for $H<H_D$ and isolated skyrmions embraced by the conical phase.
Our results are relevant not only to the application of magnetic skyrmions in memory technology, but also elucidate the fundamental properties of skyrmions and the edge states formed in the conical phases of chiral magnets.

The authors are grateful to M. Mostovoy and
T. Monchesky for useful discussions.
A.O.L acknowledges financial support by the FOM
grant 11PR2928.  A.N.B acknowledges support  
by the Deutsche Forschungsgemeinschaft 
via Grant No. BO 4160/1-1.


\begin{thebibliography} {99}

\bibitem{Leonov16a} A. O. Leonov, T. L. Monchesky, N. Romming, A. Kubetzka, A. N. Bogdanov, R. Wiesendanger, New J. of Phys. \textbf{18}, 065003 (2016).

\bibitem{Bogdanov94} A. Bogdanov and A. Hubert, J. Magn. Magn. Mater. \textbf{138}, 255 (1994).

\bibitem{Romming13} N. Romming, C. Hanneken, M. Menzel, J. E. Bickel, B. Wolter, K. von Bergmann, A. Kubetzka, R. Wiesendanger, Science \textbf{341}, 636 (2013); N. Romming, A. Kubetzka, C. Hanneken, K. von Bergmann, R. Wiesendanger, Phys. Rev. Lett. \textbf{114}, 177203 (2015).

\bibitem{Iwasaki2013} J. Iwasaki, M. Mochizuki, N. Nagaosa, 
Nat. Nanotechnol. \textbf{8}, 742 (2013). 

\bibitem{Kiselev11} N. S. Kiselev, A. N. Bogdanov, R. Schaefer, U. K. Roessler, J. Phys. D: Appl. Phys. \textbf{44}, 392001 (2011). 

\bibitem{Fert2013}A. Fert, V. Cros, J. Sampaio,  
Nat. Nanotechnol. \textbf{8}, 152 (2013). 

\bibitem{Tomasello2014}
E. M. R. Tomasello, R. Zivieri, L. Torres, M. Carpentieri, and G. Finocchio,
  Sci. Rep. \textbf{4}, 6784 (2014). 



\bibitem{Moreau2015}
C. Moreau-Luchaire, , C. Moutafis, N. Reyren, J. Sampaio, C. A. F. Vaz, N. Van Horne, K. Bouzehouane, K.
Garcia, C. Deranlot, P. Warnicke, P. Wohlh\"uter, J.-M. George, M. Weigand, J. Raabe, V. Cros, A. Fert,   Nat. Nanotechnol. \textbf{11}, 444 (2016). 

\bibitem{Woo2016} S. Woo, K. Litzius, B. Kruger, M. Y. Im, L. Caretta, K. Richter, M. Mann, A. Krone, R. M. Reeve, M. Weigand, P. Agraval, I. Lemesh, M. A. Miawass, P. Fisher, M. Klaui, G. R. S. D. Beach,  Nat. Mater. \textbf{15}, 501 (2016). 

\bibitem{Yu2012} X. Z. Yu, N. Kanazawa, W. Z. Zhang, T. Nagai, T. Hara, K. Kimoto, Y. Matsui, Y. Onose, Y. Tokura,  Nat. Commun. \textbf{3}, 988 (2012). 

\bibitem{Hanneken16} C. Hanneken, A. Kubetzka, K. von Bergmann, R. Wiesendanger, New J. Phys. \textbf{18}, 055009 (2016).

\bibitem{Du} H. F. Du, R. C. Che, L. Y. Kong, X. B. Zhao, C. M. Jin, C. Wang, J. Y. Yang, W. Ning, R. W. Li, C. Q. Jin, X. H. Chen, J. D. Zang, Y. H. Zhang, M. L. Tian, Nat. Commun. \textbf{6}, 8504 (2015); H. F. Du, J. P. DeGrave, F. Xue, D. Liang, W. Ning, J. Y. Yang, M. L. Tian, Y. H. Zhang, S. Jin, Nano Lett., \textbf{14}, 2026 (2014). 


\bibitem{twist} For an overview of this phenomenon and bibliogrphy
 see \cite{Meynell14}.

\bibitem{Meynell14} S. A. Meynell, M. N. Wilson, H. Fritzsche, A. N. Bogdanov, T. L. and Monchesky,  
Phys. Rev. B \textbf{90}, 014406 (2014). 



\bibitem{Wilson2013} M. N. Wilson,  E. A.  Karhu, D. P. Lake, A. S. Quigley, S. Meynell, A. N. Bogdanov, H. Fritzsche, U. K. R\"o\ss ler, and  T. L. Monchesky,  
Phys. Rev. B \textbf{88}, 214420 (2013). 
%
\bibitem{Keesman2015} R. Keesman, A. O.   Leonov, S.  Buhrandt, G. T.  Barkema, L. Fritz, R. A. Duine,  
Phys. Rev. B \textbf{92}, 134405 (2015). 



\bibitem{Zhang2015} X. C. Zhang, G. P. Zhao, H, Fangohr, J. P. Liu, W. X. Xia, J. Xia, F. J. Morvan, 
Sci.  Rep.  \textbf{5}, 7643 (2015).

\bibitem{Leonov16b} A. O. Leonov, T. L. Monchesky, J. C. Loudon, A N Bogdanov,  J. Phys.: Condens.
Matter. \textbf{28}, 35LT01 (2016).

\bibitem{Bak80} P. Bak P and M. H. Jensen, J. Phys.C: Solid State Phys. \textbf{13}, L881 (1980).

\bibitem{Leonov16c} A. O. Leonov, Y. Togawa, T. L. Monchesky, A. N. Bogdanov, J. Kishine, Y. Kousaka, M. Miyagawa, T. Koyama, J. Akimitsu, Ts. Koyama, K. Harada, S. Mori, D. McGrouther, R. Lamb, M. Krajnak, S. McVitie, R. L. Stamps, K. Inoue, Phys. Rev. Lett. \textbf{117}, 087202 (2016).

\end{thebibliography}
\end{document}